# Vacuum-Sealed Thermal Treatment Regulates Trap States and Red Persistent Luminescence in $CaTiO_3$: $Pr^{3+}$,$Al^{3+}$ Phosphors

Ziyi Guo[1], Duanting Yan[1*]

[1] State Key Laboratory of Integrated Optoelectronics, Key Laboratory of UV Light-Emitting Materials and Technology of Ministry of Education, School of Physics, Northeast Normal University, Changchun 130022, China.

[*]Correspongding authors

Email: [*]yandt677@nenu.edu.cn

## Abstract

Red persistent phosphors remain less mature than green and blue-green systems because their afterglow is often weak and decays rapidly. Here, $CaTiO_3$:0.3%$Pr^{3+}$,0.3%$Al^{3+}$ phosphors were prepared by a high-temperature solid-state route under either air or vacuum-sealed quartz-tube (QT) conditions. The effects of processing atmosphere and sintering temperature on phase structure, microstructure, steady-state photoluminescence, afterglow, thermoluminescence, and excited-state decay were examined. X-ray diffraction and Raman spectra show that all samples retain the orthorhombic $CaTiO_3$ perovskite phase, with no detectable secondary phase. SEM observations show particle coarsening at higher QT temperatures, while EDS mapping indicates a homogeneous distribution of Ca, Ti, O, Pr, and Al within the examined region. The QT-treated samples exhibit stronger $Pr^{3+}$ red emission near 612 nm and markedly improved afterglow compared with the air-treated sample. The QT-1300 °C sample shows the best afterglow among the present samples, with a reported 3.6-fold higher intensity at 1200 s than Air-1200 °C. Thermoluminescence results indicate that QT treatment increases the population of thermally active traps and enhances the deeper trap component. These results suggest that a low-oxygen sealed environment regulates defect-related traps, most likely involving oxygen-vacancy-associated centers, and improves carrier storage and release through the $Pr^{3+}$/$Ti^{4+}$ intervalence charge-transfer pathway. This work provides a practical processing strategy for improving $CaTiO_3$-based red persistent phosphors and offers insight into trap-state regulation under low-oxygen thermal treatment.

Keywords: $CaTiO_3$:$Pr^{3+}$,$Al^{3+}$; red persistent luminescence; vacuum-sealed treatment; trap states; thermoluminescence; oxygen vacancies

## 1. Introduction

Persistent phosphors can continue to emit light after the excitation source is removed[1-4]. This property makes them useful for safety signage, emergency lighting, anti-counterfeiting, night displays, and low-light illumination[5,6]. Green and blue-green persistent phosphors have already reached a relatively mature stage. In contrast, red persistent phosphors still face major limitations, including low brightness, fast decay, and poor control over defect traps[7,8].

$CaTiO_3$ is an attractive host for red persistent luminescence. It has a stable perovskite framework, superior chemical and thermal stability, low raw-material cost, and broad compositional tolerance [9], serving as an ideal red luminescent matrix[10,11].$Pr^{3+}$ is a common red-emitting activator in this host because its $^1D_2 \rightarrow {}^3H_4$ transition gives a narrow emission band near 612 nm[12,13]. $CaTiO_3$:$Pr^{3+}$ has become a research hotspot of red persistent luminescent materials[14,15].In $CaTiO_3$:$Pr^{3+}$, the emission and afterglow are also strongly influenced by intervalence charge transfer (IVCT) between Pr and Ti-related states. Therefore, the performance depends not only on the $Pr^{3+}$ emitting center, but also on trap states and carrier-transfer pathways in the host lattice.

Trap engineering is central to persistent luminescence. Traps that are too shallow release carriers rapidly at room temperature and cannot sustain afterglow. Traps that are too deep store carriers but cannot release them efficiently under ambient thermal activation. An optimal trap distribution should provide sufficient trap density and suitable trap depth. In oxide hosts, oxygen-vacancy-related defects often act as important carrier traps. Their concentration and distribution can be modified by oxygen partial pressure, charge compensation, and sintering temperature[16,17].

Air sintering is convenient but provides a high oxygen chemical potential. This condition may suppress oxygen deficiency and may also introduce undesired defect equilibria, depending on the composition and temperature. A vacuum-sealed quartz-tube treatment creates a low-oxygen closed environment. Such a process can favor oxygen-deficient defect formation and can help retain defect states during high-temperature treatment. When combined with $Al^{3+}$ codoping, which can modify charge compensation in the Ti sublattice, QT treatment may provide a simple route to regulate traps in $CaTiO_3:Pr^{3+}$.

In this work, $CaTiO_3:0.3\%Pr^{3+},0.3\%Al^{3+}$ powders were prepared under air and QT conditions. The QT temperature was varied from 1200 to 1300 °C. We compare phase purity, Raman vibrational features, particle morphology, elemental distribution, photoluminescence, afterglow decay, phosphorescence spectra, thermoluminescence, and luminescence decay. The goal is to clarify how QT processing and sintering temperature tune the trap landscape and thereby improve red persistent luminescence. Based on the optical and thermoluminescence results, oxygen-vacancy-associated defects are considered the most plausible origin of the trap states involved in persistent luminescence.

## 2. Experimental Section

$CaTiO_3:0.3\%Pr^{3+},0.3\%Al^{3+}$ phosphors were synthesized by a high-temperature solid-state route. $CaCO_3$, $TiO_2$, $Pr_6O_{11}$, and $Al_2O_3$ were used as starting materials. Stoichiometric amounts of the powders were dispersed in anhydrous ethanol and ground thoroughly to improve mixing. The mixed powder was dried and pre-sintered in an alumina crucible at 900 °C for 2 h. After a second grinding step, the powder was divided into two groups. One group was calcined in air at 1200 °C for 2 h and is denoted Air-1200 °C. The other group was sealed in evacuated quartz tubes and calcined at 1200, 1250, or 1300 °C for 2 h. These samples are denoted QT-1200 °C, QT-1250 °C, and QT-1300 °C, respectively.

Phase structure was examined by X-ray diffraction (XRD) using Cu Kα radiation ($\lambda$ = 1.5418 Å) over $2\theta$ = 20–80°. Particle morphology and elemental distribution were characterized by scanning electron microscopy (SEM), energy-dispersive X-ray spectroscopy (EDS), and elemental mapping. Raman spectra were collected with a 488 nm laser over 100–800 $cm^{-1}$ using a 10×objective. Excitation spectra, emission spectra, persistent luminescence decay curves, phosphorescence spectra, thermoluminescence (TL) curves, and photoluminescence decay curves were recorded at room temperature using a fluorescence spectrometer. For TL measurements, the samples were irradiated at 348 nm for 5 min and then heated at 10 K $min^{-1}$. The exact pre-irradiation, delay, and detector settings should be kept identical for all samples.

## 3. Results and discussion

Figure 1 compares the XRD patterns of the air-treated and QT-treated samples. All diffraction peaks match orthorhombic $CaTiO_3$ (PDF#22-0153). No additional peaks are observed within the detection limit. This result indicates that 0.3%$Pr^{3+}$/0.3%$Al^{3+}$ codoping and the present atmosphere/temperature range do not change the main perovskite phase. The sharp reflections of the QT-treated samples also suggest good crystallization after high-temperature treatment. However, XRD alone does not provide direct evidence for oxygen vacancies or for the detailed charge-compensation mechanism.

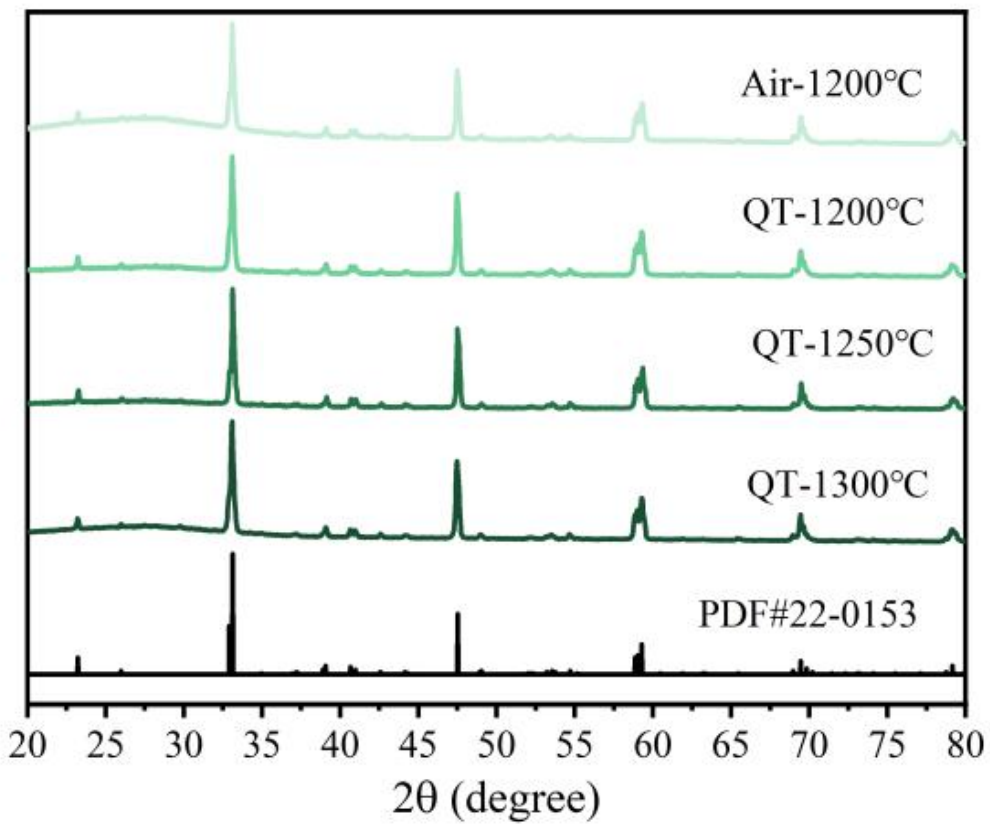

Figure 1. XRD patterns of $CaTiO_3$:0.3%$Pr^{3+}$,0.3%$Al^{3+}$ samples prepared in air at 1200 °C and by vacuum-sealed quartz-tube (QT) treatment at 1200, 1250, and 1300 °C. The standard pattern of orthorhombic $CaTiO_3$ (PDF#22-0153) is shown for comparison.

Raman spectra further support the phase assignment (Figure 2). All samples show the characteristic vibrational features of orthorhombic $CaTiO_3$. The positions of the main bands remain almost unchanged after QT treatment and after increasing the QT temperature. This indicates that the average $TiO_6$ octahedral framework remains stable. The peak intensity and sharpness become more pronounced for the QT samples, especially at higher treatment temperature, which is consistent with improved crystallinity. The current data do not show a new Raman band that can be assigned unambiguously to a secondary phase.

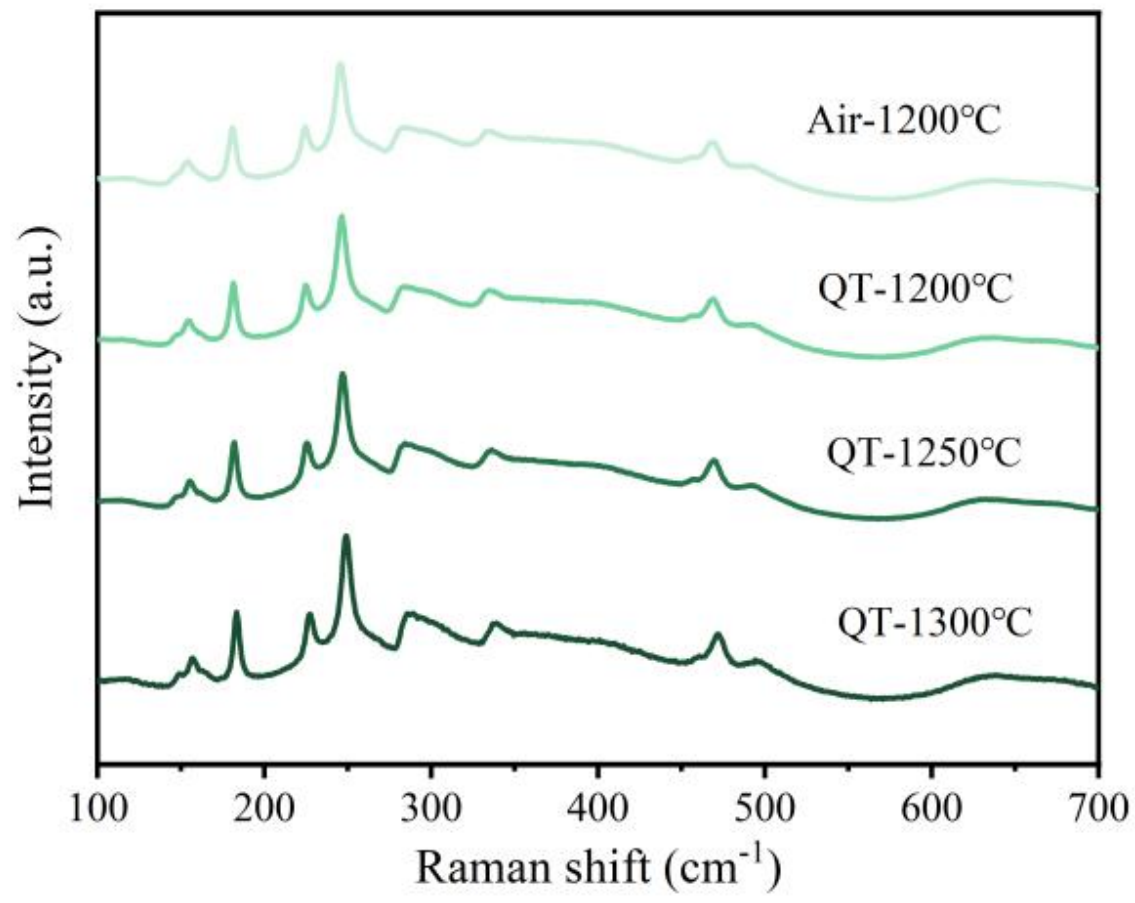

Figure 2. Raman spectra of $CaTiO_3$:0.3%$Pr^{3+}$,0.3%$Al^{3+}$ samples prepared under air and QT conditions. The spectra retain the characteristic vibrational features of orthorhombic $CaTiO_3$.

Figure 3 shows SEM images of the four samples. The powders consist of irregular and partially agglomerated particles. The air-treated and QT-treated samples at 1200 °C show broadly similar morphology. As the QT temperature increases from 1200 to 1300 °C, the particles become larger and the fraction of coarse grains increases. This behavior is expected for solid-state sintering, where higher temperature promotes diffusion, neck formation, and grain coarsening. The SEM images therefore indicate that QT temperature mainly affects particle growth rather than changing the phase composition.

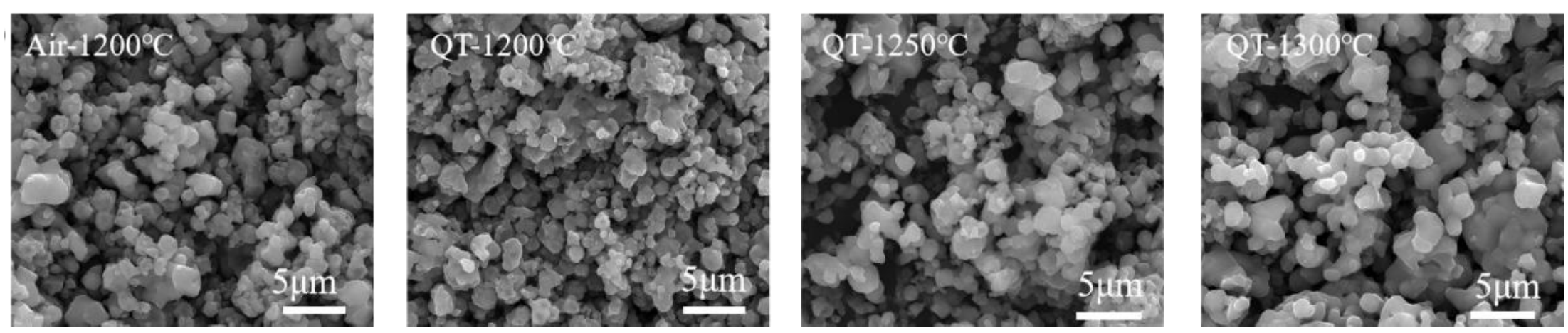

Figure 3. SEM images of $CaTiO_3$:0.3%$Pr^{3+}$,0.3%$Al^{3+}$ samples prepared in air and by QT treatment. Higher QT temperature leads to stronger particle coarsening and agglomeration.

A representative SEM/EDS mapping result is shown in Figure 4. Ca, Ti, O, Pr, and Al signals are distributed throughout the selected region. No obvious Pr-rich or Al-rich segregation is observed at the scale of the mapping image. The EDS spectrum contains Ca, Ti, O, Pr, and Al, together with Au from the conductive coating. The Ca:Ti:O ratio is close to the nominal $CaTiO_3$ stoichiometry according to the original description. Because Pr and Al are present at only 0.3%, quantitative EDS values for these dopants should be interpreted with caution. The sample identity of this mapping image should also be confirmed, because the draft contains inconsistent descriptions.

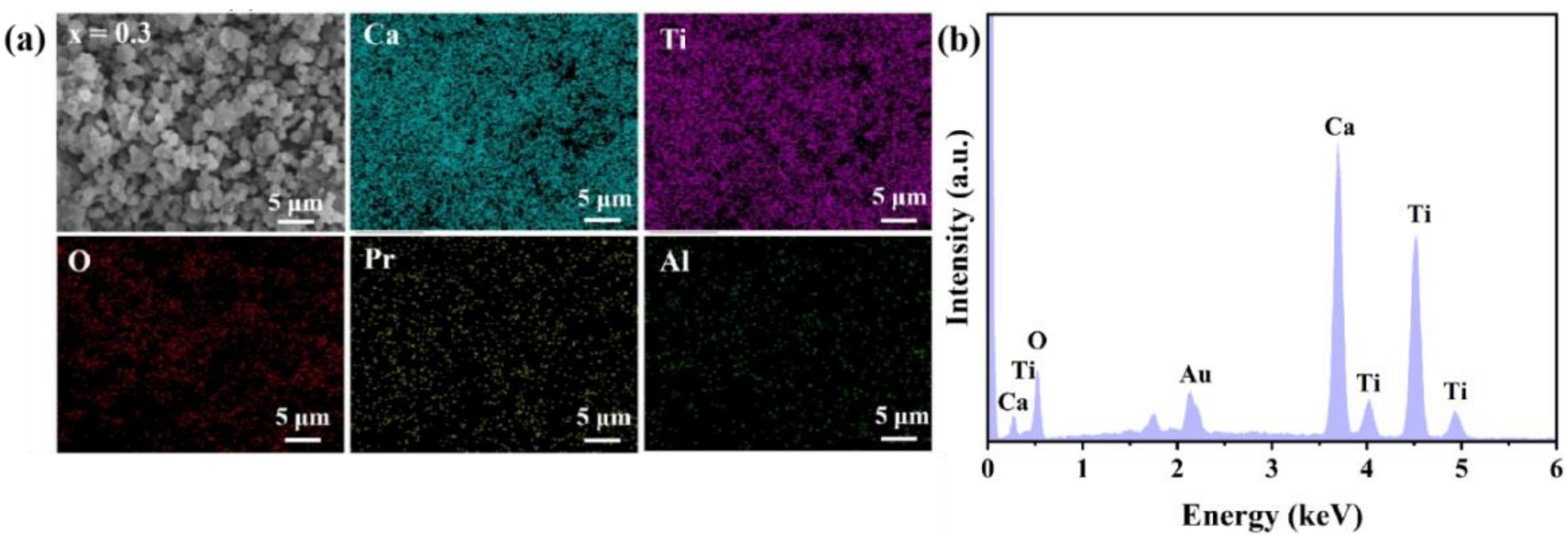

Figure 4. Representative SEM image, elemental mapping, and EDS spectrum of $CaTiO_3$:0.3%$Pr^{3+}$,0.3%$Al^{3+}$ prepared in air. The mapped region shows homogeneous distributions of Ca, Ti, O, Pr, and Al within the spatial resolution of EDS. The Au signal originates from conductive coating.

Figure 5a shows excitation spectra monitored at 612 nm. The spectra exhibit a broad UV excitation band extending from about 250 to 400 nm, with a strong maximum in the near-UV region. This band can include host absorption, defect-related absorption, and charge-transfer or IVCT-assisted excitation pathways that feed the $Pr^{3+}$ emitting state. A shoulder at longer wavelength becomes more evident in the QT samples, especially at higher temperature, suggesting that QT treatment changes the excitation pathway or the coupling between Pr-related states and defect states. Because the band is broad and contains overlapping contributions, its exact assignment requires additional excitation-resolved or temperature-dependent measurements.

Under 375 nm excitation, all samples show red emission centered near 612 nm (Figure 5b). This emission is assigned to the $Pr^{3+}$ $^1D_2 \rightarrow {}^3H_4$ transition. The emission intensity follows the order Air-1200 °C < QT-1200 °C < QT-1250 °C < QT-1300 °C. Thus, QT treatment strongly enhances the steady-state red emission, and increasing the QT temperature further improves the emission intensity. This enhancement is consistent with improved crystallization and modified defect states. However, PL intensity alone cannot distinguish whether the dominant effect is stronger absorption, more efficient energy transfer, reduced nonradiative loss, or a higher number of active Pr centers.

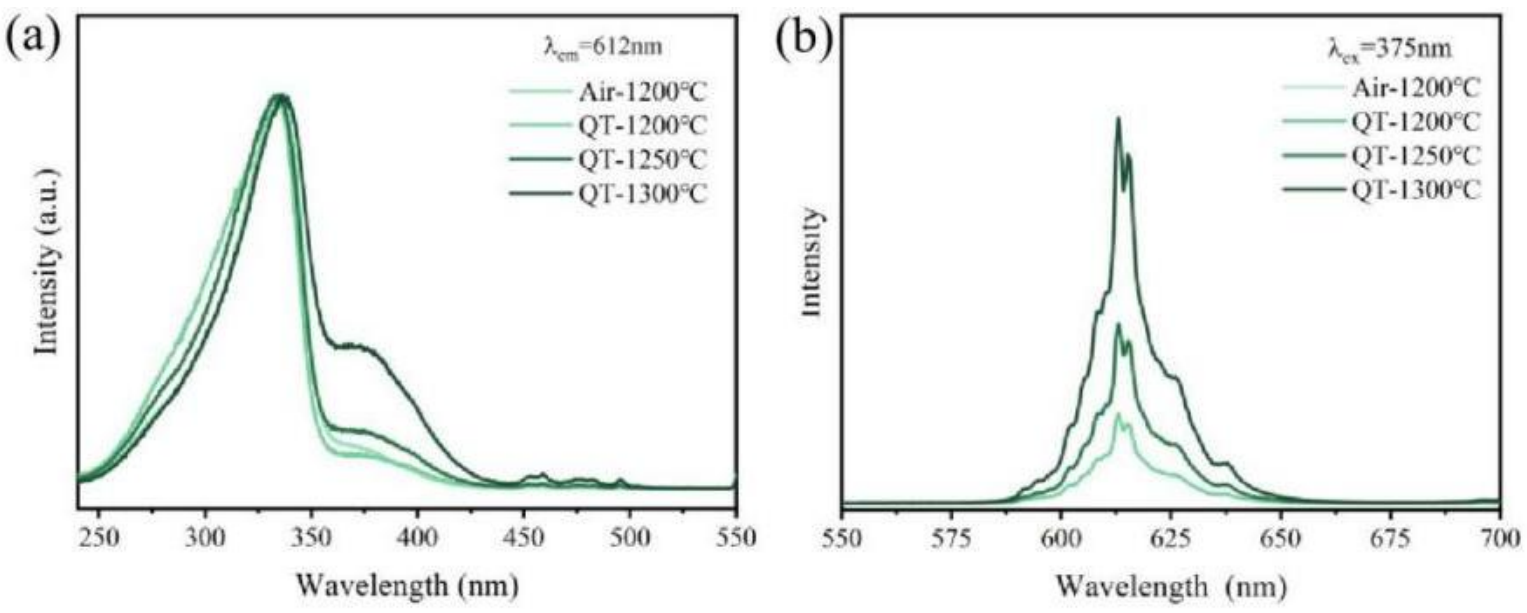

Figure 5. Photoluminescence spectra of $CaTiO_3$:0.3%$Pr^{3+}$,0.3%$Al^{3+}$ samples: (a) excitation spectra monitored at 612 nm and (b) emission spectra recorded under 375 nm excitation.

The afterglow decay curves monitored at 612 nm are shown in Figure 6a. All samples display a rapid initial decay followed by a slower long-time tail. This behavior is typical of persistent phosphors with traps of different release rates. The air-treated sample decays fastest and maintains the weakest long-time signal. QT treatment clearly raises the afterglow intensity. Among the QT samples, the long-time afterglow improves as the treatment temperature increases, and QT-1300 °C gives the strongest signal over the measured time window. The original draft reports that the afterglow intensity at 1200 s is 3.6 times higher for the optimized QT sample than for Air-1200 °C.

The phosphorescence spectra recorded after UV irradiation (Figure 6b) show the same emission band near 612 nm for all samples. Their spectral shapes are similar, whereas their intensities differ strongly. This result confirms that the persistent luminescence and the steady-state PL originate from the same $Pr^{3+}$ red-emitting center. The improvement in afterglow therefore mainly reflects changes in carrier storage and release rather than the formation of a new emitting center.

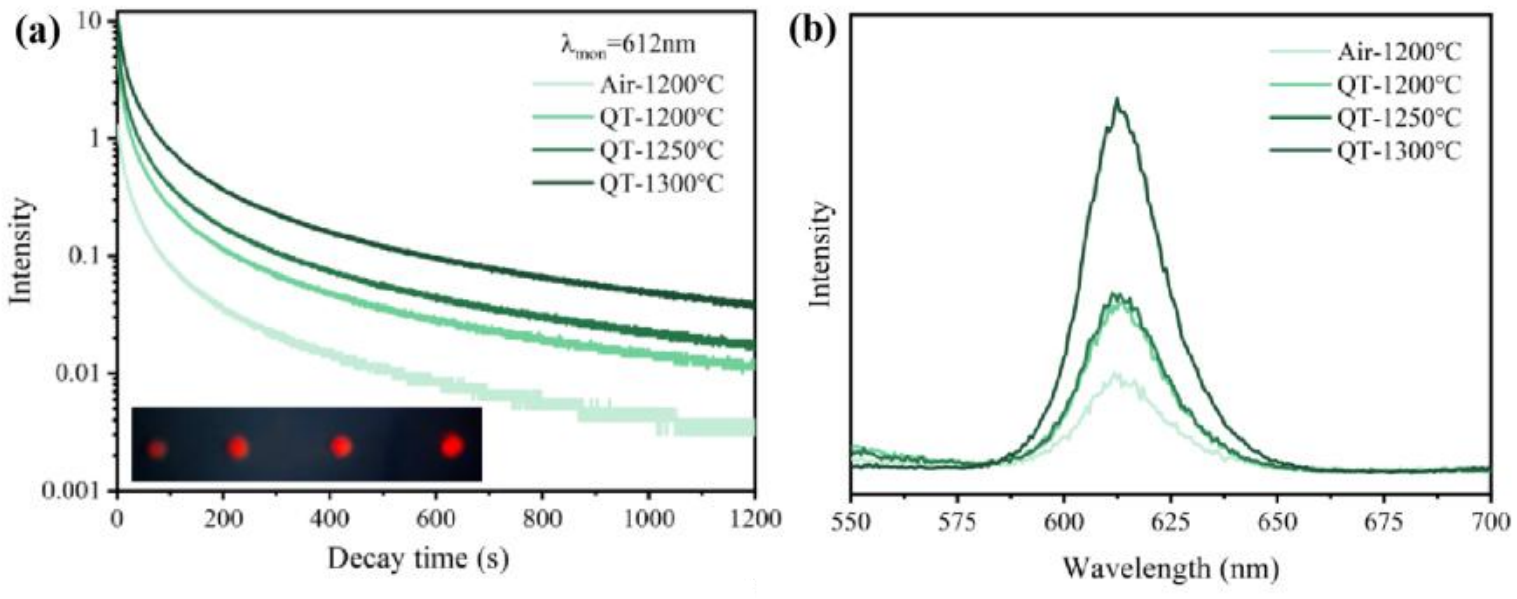

Figure 6. Persistent luminescence of $CaTiO_3$:0.3%$Pr^{3+}$,0.3%$Al^{3+}$ samples. (a) Afterglow decay curves monitored at 612 nm after UV irradiation. The inset shows photographs taken after the excitation was stopped. (b) Phosphorescence spectra recorded after 348 nm irradiation.

TL glow curves provide information on thermally activated carrier release from traps (Figure 7). The QT samples show much stronger TL signals than Air-1200 °C. This indicates that QT treatment increases the concentration of traps that can release carriers near and above room temperature. With increasing QT temperature, the TL intensity increases further, and the broad signal extends more clearly toward higher temperature. This trend suggests that high-temperature QT treatment increases the number of effective traps and may also enhance a deeper trap component.

The present TL curves support the trap-regulation mechanism, but they do not identify the chemical nature of the traps by themselves. A low-oxygen sealed environment and $Al^{3+}$ codoping both make oxygen-vacancy-associated defects plausible. These defects may capture carriers and then release them thermally to feed the $Pr^{3+}$ red-emitting state. Direct evidence from EPR, XPS, positron annihilation, or controlled oxygen partial pressure experiments would be needed to confirm this assignment. Quantitative trap depths also require glow-curve deconvolution or variable heating-rate analysis.

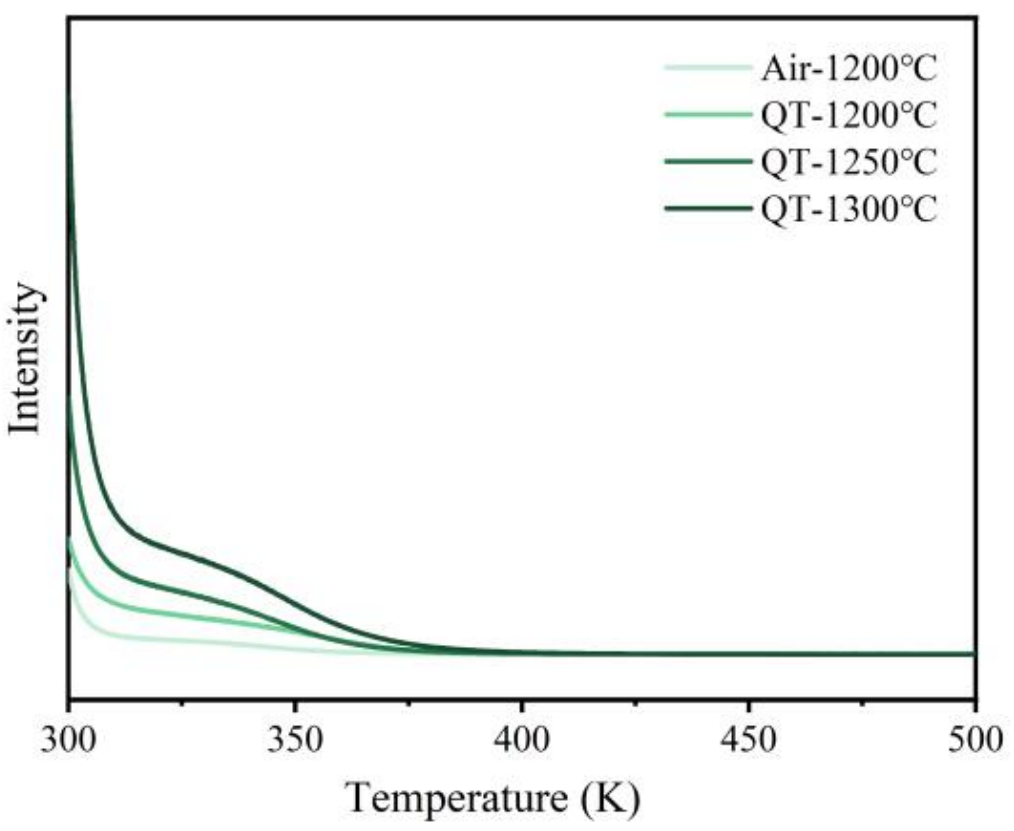

Figure 7. Thermoluminescence glow curves of $CaTiO_3$:0.3%$Pr^{3+}$,0.3%$Al^{3+}$ samples after 348 nm irradiation for 5 min. The heating rate was 10 K $min^{-1}$.

Figure 8 shows normalized PL decay curves monitored at 612 nm under 348 nm excitation. The decay is not a simple single-exponential process, which indicates that the $Pr^{3+}$ excited state is coupled to more than one relaxation pathway. The air-treated sample appears to show the slowest normalized decay. QT treatment shortens the decay relative to Air-1200 °C, but the trend among QT samples is not simply monotonic. In the displayed curves, QT-1200 °C decays fastest, whereas QT-1300 °C shows a somewhat slower decay than QT-1200 °C and QT-1250 °C.

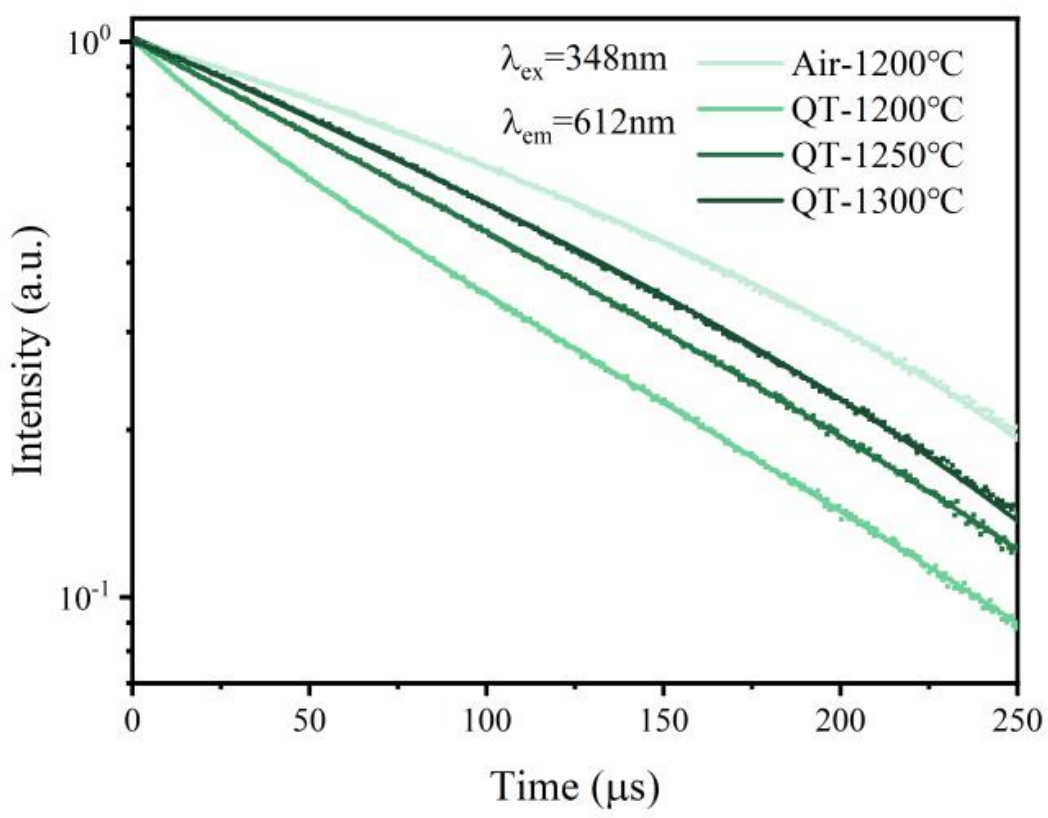

Figure 8. Normalized photoluminescence decay curves of $CaTiO_3$:0.3%$Pr^{3+}$,0.3%$Al^{3+}$ samples monitored at 612 nm under 348 nm excitation.

The decay behavior indicates that QT treatment modifies the excited-state relaxation dynamics of the $Pr^{3+}$ center. A stronger steady-state emission can arise from enhanced absorption, improved energy transfer, and increased radiative efficiency. The non-single-exponential decay behavior suggests the coexistence of multiple relaxation pathways associated with defect-assisted carrier transfer. These results indicate that QT processing significantly influences the excited-state dynamics of the $Pr^{3+}$ center.

Figure 9 summarizes a possible mechanism for red persistent luminescence in $CaTiO_3:Pr^{3+},Al^{3+}$. The $Pr^{3+}$ $^1D_2$ state is the main emitting level and relaxes radiatively to $^3H_4$, producing the 612 nm red emission. In Pr-doped titanates, the $Pr^{3+}/Ti^{4+}$ IVCT state can mediate energy transfer and can compete with the higher-lying $^3P_J$ levels of $Pr^{3+}$. This pathway helps populate $^1D_2$ and explains why red emission dominates.

Under UV or near-UV excitation, carriers can be generated through host absorption, IVCT-assisted excitation, or defect-related absorption. A fraction of these carriers is captured by traps, most likely oxygen-vacancy-associated centers or related defect complexes. After the excitation source is removed, trapped carriers are gradually released by thermal activation. They then recombine through the Pr/Ti-related IVCT pathway and populate the $Pr^{3+}$ $^1D_2$ state, producing delayed red emission. The afterglow intensity depends on trap density, trap depth, and carrier-transfer efficiency.

QT treatment likely improves the afterglow because it favors the formation or retention of oxygen-deficient traps. Increasing the QT temperature promotes crystallization and increases the effective trap population, as supported by the stronger TL signals. The optimized QT-1300 °C sample therefore combines stronger red emission with more efficient long-time carrier storage and release. This mechanism is consistent with the XRD, Raman, SEM, PL, afterglow, TL, and decay results, but the detailed chemical identity and concentration of the traps still require direct defect characterization.

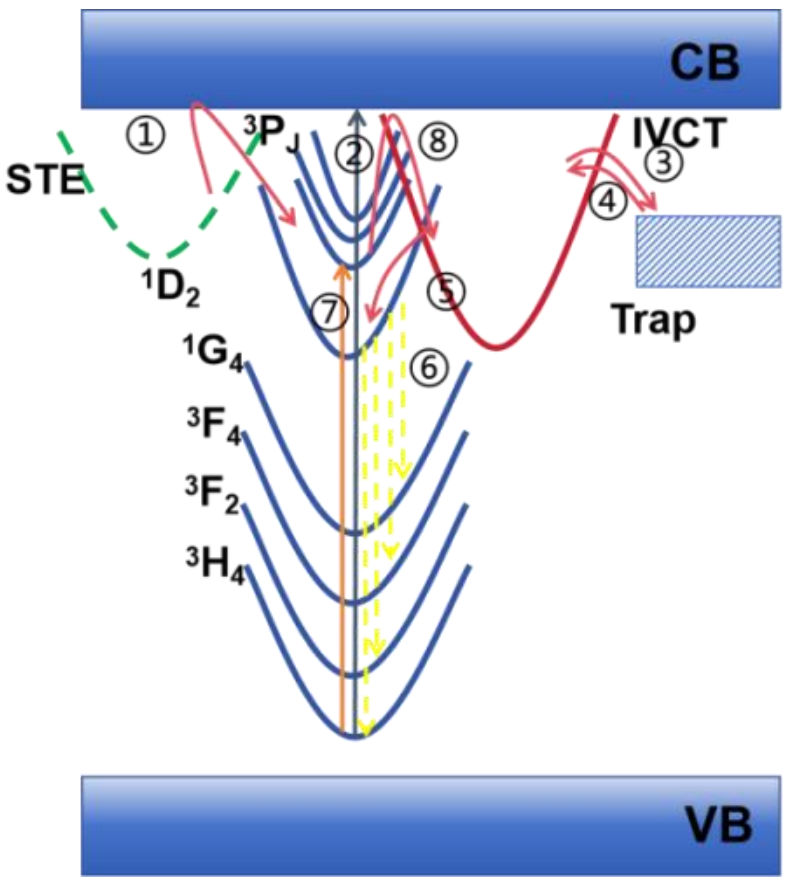

Figure 9. Schematic illustration of the possible persistent-luminescence mechanism in $CaTiO_3:Pr^{3+},Al^{3+}$. $Pr^{3+}$ emission, IVCT-mediated transfer, and defect-related traps jointly determine carrier storage, release, and delayed red emission.

## 4. Conclusions

$CaTiO_3$:0.3%$Pr^{3+}$,0.3%$Al^{3+}$ red persistent phosphors were prepared by solid-state reaction under air and vacuum-sealed quartz-tube conditions. All samples retain the orthorhombic $CaTiO_3$ phase, and no secondary phase is detected by XRD or Raman spectroscopy. QT treatment enhances the $Pr^{3+}$ red emission near 612 nm and strongly improves persistent luminescence. The QT-1300 °C sample shows the best afterglow among the studied samples, with a reported 3.6-fold higher intensity at 1200 s than the air-treated sample. Thermoluminescence results show that QT treatment increases the number of thermally active traps and that higher QT temperature further strengthens the effective trap component. These findings suggest that a low-oxygen sealed environment regulates oxygen-vacancy-associated traps and improves carrier storage/release through Pr/Ti-related IVCT pathways. The results demonstrate that vacuum-sealed thermal treatment is a useful processing route for improving $CaTiO_3$-based red persistent phosphors.These results demonstrate that vacuum-sealed thermal treatment is an effective strategy for regulating trap states and enhancing red persistent luminescence in $CaTiO_3$-based phosphors.